\documentclass{ws-procs975x65}

\begin{document}

\title{Scalar and Pseudoscalar Glueballs}

\author{Hai-Yang Cheng$^*$}

\address{Institute of Physics, Academia Sinica,\\
Taipei, Taiwan 115, ROC\\
$^*$E-mail: phcheng@phys.sinica.edu.tw\\
}

\begin{abstract}
We employ two simple and robust results to constrain the mixing matrix of the isosinglet scalar mesons
$f_0(1710)$, $f_0(1500)$, $f_0(1370)$: one is the approximate SU(3) symmetry empirically observed in the scalar sector
above 1 GeV and confirmed by lattice QCD, and the other is the scalar glueball mass at 1710
MeV in the quenched approximation. In the SU(3) symmetry limit,
$f_0(1500)$ becomes a pure SU(3) octet and is degenerate with
$a_0(1450)$, while $f_0(1370)$ is mainly an SU(3) singlet with a
slight mixing with the scalar glueball which is the primary
component of $f_0(1710)$. These features remain essentially
unchanged even when SU(3) breaking is taken into account.  The
observed enhancement of $\omega f_0(1710)$ production over $\phi
f_0(1710)$ in hadronic $J/\psi$ decays and the copious $f_0(1710)$
production in radiative $J/\psi$ decays lend further support to
the prominent glueball nature of $f_0(1710)$. We deduce the mass of the pseudoscalar glueball $G$ from an
$\eta$-$\eta'$-$G$ mixing formalism based on the anomalous Ward
identity for transition matrix elements. With the inputs from the
recent KLOE experiment, we find a solution for the pseudoscalar
glueball mass around $(1.4\pm 0.1)$ GeV, which is fairly insensitive
to a range of inputs with or without Okubo-Zweig-Iizuka-rule
violating effects. This affirms that $\eta(1405)$, having a large
production rate in the radiative $J/\psi$ decay and not seen in
$\gamma\gamma$ reactions, is indeed a leading candidate for the
pseudoscalar glueball. It is much lower than the results
from quenched lattice QCD ($>2.0$ GeV) due to the dynamic fermion effect. It is thus urgent to have a full QCD lattice calculation of pseudoscalar glueball masses.

\end{abstract}


\bodymatter

\section{Introduction}\label{aba:sec1}
The existence of glueballs is an archetypal prediction of QCD as a
confining theory. Despite a great deal of experimental and
theoretical efforts in the last 3 decades, there is still no any compelling evidence for their existence.
Surely, there are good
glueball candidates, such as $f_0(1710)$ and $\eta(1405)$ which
are seen copioursly in $J/\psi$ radiative decays and yet not
observed in $\gamma\gamma$ reactions~\cite{PDG}. However,
definitive identification has been plagued by the complication
that the branching ratios of these candidates in the radiative decay
of $J/\psi$, once thought to be the defining channel to probe the
glue-rich content of the mesons, are not orders of magnitude large
than the other known $q\bar{q}$ mesons, that the glueball can mix
with ordinary mesons, and the fact that the statistics in
experiments are neither high enough to confidently detect all the
major decay channels nor precise enough to disentangle from the
near-by states such as $f_0(1790)$ and $\eta(1475)$. In this case,
the supplemental information on the quark content (or rather the
lack of) via the $\gamma\gamma$ coupling and leptonic decays prove
to be important to reveal the nature of the glueball candidates.

\section{Scalar glueball}
It is generally believed that the scalar glueball is hidden itself somewhere in the isosinglet scalar mesons with masses above 1 GeV. The argument goes as follows. Many scalar mesons with masses lower than 2 GeV have been observed and they can be classified into two nonets: one nonet with mass below or close to 1 GeV, such as $\sigma,~\kappa$, $f_0(980)$ and $a_0(980)$ that are generally believed to be composed of four quarks and the other nonet with mass above 1 GeV such as $K_0^*(1430)$, $a_0(1450)$ and two isosinglet scalar mesons. This means that not all three isosinglet scalars $f_0(1710)$, $f_0(1500)$, $f_0(1370)$ can be accommodated in the $q\bar q$ nonet picture. One of them should be primarily a scalar glueball.

Among the isosinglet scalar mesons $f_0(1710)$, $f_0(1500)$ and
$f_0(1370)$, it has been quite controversial as to which of these
is the dominant scalar glueball. It has been suggested that
$f_0(1500)$ is primarily a scalar glueball \cite{Close1}, due
partly to the fact that $f_0(1500)$, discovered in $p\bar{p}$
annihilation at LEAR, has decays to $\eta\eta$ and $\eta\eta'$
which are relatively large compared to that of
$\pi\pi$~\cite{ams95} and that the earlier quenched lattice
calculations~\cite{bsh93} predict the scalar glueball mass to be
around $1550$ MeV. Furthermore, because of the small production of
$\pi\pi$ in $f_0(1710)$ decay compared to that of $K\bar K$, it
has been thought that $f_0(1710)$ is primarily $s\bar s$
dominated. In contrast, the smaller production rate of $K\bar K$
relative to $\pi\pi$ in $f_0(1370)$ decay leads to the conjecture
that $f_0(1370)$ is governed by the non-strange light quark
content.

Based on the above observations, a flavor-mixing scheme is
proposed \cite{Close1} to consider the glueball and $q\bar q$
mixing in the neutral scalar mesons $f_0(1710)$, $f_0(1500)$ and
$f_0(1370)$.  Best $\chi^2$ fits to the measured scalar meson
masses and their branching ratios of strong decays have been
performed in several references by Amsler, Close and
Kirk~\cite{Close1}, Close and Zhao~\cite{Close2}, and He {\it et
al.}~\cite{He}. A typical mixing matrix in this scheme is
\cite{Close2}
 \begin{eqnarray} \label{eq:Close}
 \left(\begin{matrix} f_0(1370) \cr f_0(1500) \cr f_0(1710) \cr\end{matrix}\right)=
\left( \begin{matrix} -0.91 & -0.07 & 0.40 \cr
                 -0.41 & 0.35 & -0.84 \cr
                0.09 & 0.93 & 0.36 \cr
                  \end{matrix}\right)\left(\begin{matrix}|N\rangle \cr
 |S\rangle \cr |G\rangle \cr\end{matrix}\right). \nonumber
 \end{eqnarray}
A common feature of these analyses is that, before mixing, the
$s\bar{s}$ mass $M_S$ is larger than the glueball mass $M_G$
which, in turn, is larger than the
$N(\equiv(u\bar{u}+d\bar{d})/\sqrt{2})$ mass $M_N$, with $M_G$
close to 1500 MeV and $M_S-M_N$ of the order of $200\sim 300$ MeV.

Other scenarios also have been proposed. For example, based on their lattice
calculations of the quenched scalar glueball mass, Lee and
Weingarten~\cite{lw97,Lee} considered a mixing scheme where
$f_0(1500)$ is an almost pure $s\bar{s}$ meson and $f_0(1710)$ and
$f_0(1370)$ are primarily the glueball and $u\bar{u}+d\bar{d}$
meson respectively, but with substantial mixing between the two
($\sim 25\%$ for the small component). With the effective chiral
Lagrangian approach, Giacosa {\it et al.}~\cite{Giacosa} performed
a fit to the experimental masses and decay widths of $f_0(1710)$,
$f_0(1500)$ and $f_0(1370)$ and found four possible solutions,
depending on whether the direct decay of the glueball component is
considered. In spit of different opinions in the community on the identification of the scalar glueball, Particle Data Group \cite{PDG06} tried to conclude the status as ``Experimental evidence is mounting that $f_0(1500)$ has considerable affinity for glue and that the $f_0(1370)$ and $f_0(1710)$ have large $u\bar u+d\bar d$ and $s\bar s$ components, respectively".

However, there are at least four serious problems with the above
scenario: (i) The isovector scalar meson $a_0(1450)$ is now
confirmed to be the $q\bar{q}$ meson in the lattice
calculation~\cite{Mathur}. As such, the degeneracy of $a_0(1450)$
and $K_0^*(1430)$, which has a strange quark, cannot be explained
if $M_S$ is larger than $M_N$ by $\sim 250$ MeV. (ii) The most
recent quenched lattice calculation with improved action and
lattice spacings extrapolated to the continuum favors a larger
scalar glueball mass close to 1700 MeV~\cite{Chen,MP}. (iii) If
$f_0(1710)$ is dominated by the $s\bar s$ content, the decay
$J/\psi\to \phi f_0(1710)$ is expected to have a rate larger than
that of $J/\psi\to \omega f_0(1710)$. Experimentally, it is other
way around: the rate for $\omega f_0(1710)$ production is about 6
times that of $J/\psi\to \phi f_0(1710)$. (iv) It is well known
that the radiative decay $J/\psi\to \gamma f_0$ is an ideal place
to test the glueball content of $f_0$. If $f_0(1500)$ has the
largest scalar glueball component, one expects  the
$\Gamma(J/\psi\to \gamma f_0(1500))$ decay rate to be
substantially larger than that of $\Gamma(J/\psi\to \gamma
f_0(1710))$. Again, experimentally, the opposite is true. Simply
based on the above experimental observations, one will naively
expect that $\gamma\gg\alpha>\beta$ in the wave function of
$|f_0(1710)\rangle=\alpha|N\rangle+\beta|S\rangle+\gamma|G\rangle$.

In our recent work \cite{CCL}, we have employed two simple and robust
results as the input for the mass matrix which is essentially the
starting point for the mixing model between scalar mesons and the
glueball. First of all, we know empirically that flavor SU(3) is an approximate symmetry in the scalar meson sector above 1 GeV. The near degeneracy of $K_0^*(1430)$, $a_0(1470)$, and $f_0(1500)$ has been observed. In the scalar charmed meson sector, $D_{s0}^*(2317)$ and $D_0^*(2308)$ have similar masses even though the former contains a strange quark. It is most likely that the same phenomenon also holds in the scalar bottom meson sector. This feature is also confirmed by the quenched lattice
calculation of the masses for the isovector scalar meson $a_0$ and the axial-vector meson $a_1$ \cite{Mathur}. It
is found that, when the quark mass is smaller than that of the
strange, $a_0$ mass is almost independent of the quark mass, in
contrast to those of $a_1$ and other hadrons that have been
calculated on the lattice (see Fig. 1).  This explains the fact that
$K_0^{*}(1430)$ is basically degenerate with $a_0(1450)$ despite
having one strange quark. This unusual behavior is not understood
as far as we know and it serves as a challenge to the existing
hadronic models. In any case, these lattice results hint at an
SU(3) symmetry in the scalar meson sector.
Second, an improved quenched lattice calculation
of the glueball spectrum at the infinite volume and continuum
limits based on much larger and finer lattices have been carried
out~\cite{Chen}. The mass of the scalar glueball is calculated to
be $m(0^{++})=1710\pm50\pm 80$ MeV.   This suggests that $M_G$
should be close to 1700 MeV rather than 1500 MeV from the earlier
lattice calculations~\cite{bsh93}.

\begin{figure}[t]
\centerline{\psfig{file=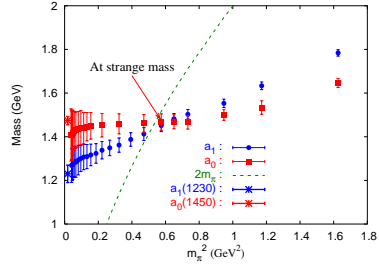,width=2.0in}} \caption{Lattice
calculations of $a_0$ and $a_1$ masses as a function of $m_\pi^2$
[6].} \label{fig1}
\end{figure}

 We shall use $|U\rangle, |D\rangle, |S\rangle$ to denote the
quarkonium states $|u\bar u\rangle, |d\bar d\rangle$ and $|s\bar s\rangle$,
and $|G\rangle$ to denote the pure scalar glueball state. In this
basis, the mass matrix reads
\begin{eqnarray} \label{eq:massmatrix}
 {\rm M}=\left( \begin{matrix} M_U & 0 & 0 & 0  \cr
                   0 & M_D & 0 & 0  \cr
                    0 & 0 & M_S & 0  \cr
                      0 & 0 & 0 & M_G  \cr
                  \end{matrix} \right)+\left(\begin{matrix} x & x & x_s & y \cr
                x & x & x_s & y \cr
                x_s & x_s & x_{ss} & y_s \cr
                y & y & y_s & 0 \cr\end{matrix} \right),
\end{eqnarray}
where the parameter $x$ denotes the mixing between different
$q\bar{q}$ states through quark-antiquark annihilation and $y$
stands for the glueball-quarkonia mixing strength. Possible SU(3)
breaking effects are characterized by the subscripts ``$s$" and
``$ss$". As noticed in passing, lattice calculations~\cite{Mathur}
of the $a_0(1450)$ and $K_0^*(1430)$ masses indicate a good SU(3)
symmetry for the scalar meson sector above 1 GeV. This means that
$M_S$ should be close to $M_U$ and $M_D$. Also the glueball mass
$M_G$ should be close to the scalar glueball mass $1710\pm50\pm80$
MeV from the lattice QCD calculation in the pure gauge sector
\cite{Chen}.

We begin by considering exact SU(3) symmetry as a first
approximation, namely, $M_S=M_U=M_D=M$ and $x_s=x_{ss}=x$ and
$y_s=y$. In this case, two of the mass eigenstates are identified
with $a_0(1450)$ and $f_0(1500)$ which are degenerate with the
mass $M$. Taking $M$ to be the experimental mass of $1474\pm 19$
MeV \cite{PDG}, it is a good approximation for the mass of
$f_0(1500)$ at $1507\pm 5$ MeV \cite{PDG}. Thus, in the limit of
exact SU(3) symmetry, $f_0(1500)$ is the SU(3) isosinglet octet
state $|f_{\rm octet}\rangle$ and is degenerate with $a_0(1450)$. In the absence of glueball-quarkonium mixing, i.e. $y=0$, $f_0(1370)$
becomes a pure SU(3) singlet $|f_{\rm singlet}\rangle$ and $f_0(1710)$
the pure glueball $|G\rangle$. The $f_0(1370)$ mass is given by
$m_{f_0(1370)}=M+3x$. Taking the experimental $f_0(1370)$ mass to
be $1370$ MeV,  $x$ is found to be $-33$ MeV.  When the
glueball-quarkonium mixing $y$ is turned on, there will be some
mixing between the glueball and the SU(3)-singlet $q\bar{q}$ . If
$y$ has the same magnitude as $x$,  the mass
shift of $f_0(1370)$ and $f_0(1710)$ due to mixing is only of order 10 MeV, a feature confirmed by the lattice calculation \cite{Lee}.

 As discussed before, SU(3) symmetry leads naturally to the near
degeneracy of $a_0(1450)$, $K_0^*(1430)$ and $f_0(1500)$. However,
in order to accommodate the observed branching ratios of strong
decays, SU(3) symmetry must be broken to certain degree in the
mass matrix and/or in the decay amplitudes. One also needs
$M_S>M_U=M_D$ in order to lift the degeneracy of $a_0(1450)$ and
$f_0(1500)$. Since the SU(3) breaking effect is expected to be
weak, they will be treated perturbatively.

If $f_0(1710)$ is primarily a pseudoscalar glueball, it is naively expected that $\Gamma(G\to\pi\pi)/\Gamma(G\to K\bar K)\approx 0.9$ after phase space correction due to the flavor independent coupling of $G$ to $PP$. However, experimentally there is a relatively large suppression of $\pi\pi$ production
relative to $K\bar K$ in $f_0(1710)$ decay.
To explain the large disparity between $\pi\pi$ and $K\bar K$
production in scalar glueball decays, Chanowitz \cite{Chanowitz}
advocated that a pure scalar glueball cannot decay into
quark-antiquark in the chiral limit, i.e.
$A(G\to q\bar q)\propto m_q$.
Since the current strange quark mass is an order of magnitude
larger than $m_u$ and $m_d$, decay to $K\bar K$ is largely favored
over $\pi\pi$. Furthermore, it has been pointed out that chiral
suppression will manifest itself at the hadron level~\cite{Chao}.
To this end, it has been suggested \cite{Chao} that $m_q$  should be interpreted as the scale of
chiral symmetry breaking since chiral symmetry is broken not only
by finite quark masses but is also broken spontaneously.
Consequently, chiral suppression for the ratio $\Gamma(G\to
\pi\pi)/\Gamma(G\to K\bar K)$ is not so strong as the current
quark mass ratio $m_u/m_s$.

Guided by the lattice calculations for chiral suppression in $G\to
PP$ \cite{Sexton}, we have performed a best $\chi^2$ fit to the measured masses and branching ratios. The mixing matrix obtained in our model has
the form:
\begin{eqnarray}  \label{eq:wf}
 \left(\begin{matrix} f_0(1370) \cr f_0(1500) \cr f_0(1710) \cr\end{matrix}\right)=
\left( \begin{matrix} 0.78 & 0.51 & -0.36 \cr
                 -0.54 & 0.84 & 0.03 \cr
                0.32 & 0.18 & 0.93 \cr
                  \end{matrix}\right)\left(\begin{matrix}|N\rangle \cr
 |S\rangle \cr |G\rangle \cr\end{matrix}\right).
\end{eqnarray}
It is evident that $f_0(1710)$ is composed primarily of the scalar
glueball, $f_0(1500)$ is close to an SU(3) octet, and $f_0(1370)$
consists of an approximate SU(3) singlet with some glueball
component ($\sim 10\%$). Unlike $f_0(1370)$, the glueball content
of $f_0(1500)$ is very tiny because an SU(3) octet does not mix
with the scalar glueball. Because the $n\bar
n$ content is more copious than $s\bar s$ in $f_0(1710)$, it is
natural that $J/\psi\to\omega f_0(1710)$ has a rate larger than
$J/\psi\to \phi f_0(1710)$. Our prediction of $\Gamma(J/\psi\to
\omega f_0(1710))/\Gamma(J/\psi\to \phi f_0(1710))=4.1$ is
consistent with the observed value of $6.6\pm2.7$. Moreover,  if $f_0(1710)$ is composed mainly
of the scalar glueball, it should be the most prominent scalar
produced in the radiative $J/\psi$ decay. Hence, it is expected that
$\Gamma(J/\psi\to \gamma f_0(1710))\gg \Gamma(J/\psi\to \gamma
f_0(1500))$, a relation borne out by experiment. Finally, we remark that in our mixing model, the relative $2\gamma$
coupling strength is $f_0(1370):f_0(1500):f_0(1710)=9.3:1.0:1.5$.
Hence $f_0(1500)$ has the smallest $2\gamma$ coupling of the three
states even though it has the least glue content in our model. Therefore, the fact that $f_0(1500)$ has not been seen in $\gamma\gamma$ reactions doesn't necessarily imply its glueball content.

\section{Pseudoscalar glueball}
In 1980, Mark II observed  a
resonance with a mass around 1440 MeV in the radiative $J/\psi$ decay \cite{iota} and identified it with the $E(1420)$ meson first discovered at CERN in 1963
through $p\bar p$ interactions \cite{E(1420)}. It was then realized that the new state observed by Mark II was not $E(1420)$ and was subsequently named $\iota(1440)$ by Mark II and Crystal Ball Collaborations \cite{CB}.
Shortly after the Mark II experiment, $\iota(1440)$ now known as
$\eta(1405)$
was proposed to be a leading candidate for the pseudoscalar glueball. (For an excellent review of the $E$ and $\iota$ mesons,
see \cite{MCU06}.) Indeed $\eta(1405)$ behaves like a glueball in its productions and
decays because it has a large
production rate in the radiative $J/\psi$ decay and is not seen in
$\gamma\gamma$ reactions. Besides $\eta(1405)$, other states
with masses below 2 GeV have also been proposed as the candidates,
such as $\eta(1760)$  and $X(1835)$.

However, the pseudoscalar glueball interpretation
for $\eta(1405)$ is not favored by most of the theoretical calculations. For example,
quenched lattice gauge
calculations predict the mass of the $0^{-+}$ state to be
above 2 GeV in \cite{bsh93} and around 2.6 GeV in
\cite{Morningstar,Chen}. It is not favored by the sum-rule analysis
with predictions higher than 1.8 GeV \cite{sum,GG97} either. Hence, we are encountering an embarrassing situation that although experimentally $\eta(1405)$ is a favored candidate for the pseudoscalar glueball, theorists seem to prefer to have a
$0^{-+}$ state heavier than the scalar glueball. The motivation of our recent work \cite{CLL} is to see if we can learn something about the glueball mass by studying the $\eta-\eta'-G$ mixing.

The $\eta-\eta'$ mixing has been well studied by Feldmann, Kroll and Stech \cite{FKS}. We extend the FKS formalism
to include the pseudoscalar glueball $G$. In the FKS scheme, the
conventional singlet-octet basis and the quark-flavor basis have
been proposed. For the latter, the $q\bar q\equiv (u\bar u+d\bar
d)/\sqrt{2}$ and $s\bar s$ flavor states, labeled by the $\eta_q$
and $\eta_s$ mesons, respectively, are defined. The physical states $\eta$, $\eta'$
and $G$ are related to the octet, singlet, and unmixed glueball
states $\eta_8$, $\eta_1$ and $g$, respectively, through the
combination of rotations
\begin{eqnarray}\label{qs}
   \left( \begin{array}{c}
    |\eta\rangle \\ |\eta'\rangle\\|G\rangle
   \end{array} \right)
   &=& \left( \begin{matrix} \cos\phi+\sqrt{2/3}\sin\theta\Delta_G & -\sin\phi+\sqrt{1/3}\sin\theta\Delta_G & -\sin\theta \sin\phi_G \cr
   \sin\phi-\sqrt{2/3}\sin\theta\Delta_G & \cos\phi-\sqrt{1/3}\cos\theta\Delta_G & \cos\theta \sin\phi_G \cr
   -\sqrt{2/3}\sin\phi_G & -\sqrt{1/3}\sin\phi_G & \cos\phi_G
   \end{matrix} \right)
   \left( \begin{array}{c}
    |\eta_8\rangle \\ |\eta_1\rangle\\|g\rangle
   \end{array} \right)   \nonumber \\
   &\equiv& U(\phi,\phi_G) \left( \begin{array}{c}
    |\eta_8\rangle \\ |\eta_1\rangle\\|g\rangle
   \end{array} \right) \;,
\end{eqnarray}
where $\theta$ is the $\eta-\eta'$ mixing angle in the octet-singlet basis, $\phi=\theta+54.7^\circ$, $\Delta_G=1-\cos\phi_G$ and $\phi_G$ is the mixing angle between $G$ and $\eta_1$; that is, we have assumed that $\eta_8$ does not mix with the glueball.

We proceed to define decay constants for the physical and flavor states
\begin{eqnarray}
   \langle 0|\bar q\gamma^\mu\gamma_5 q|\eta_q(P)\rangle
   = -\frac{i}{\sqrt2}\,f_q\,P^\mu,    & &\langle 0|\bar q\gamma^\mu\gamma_5 q|\eta_s(P),g(P)\rangle
   = -\frac{i}{\sqrt2}\,f_{s,g}^q\,P^\mu \;, \nonumber\\
   \langle 0|\bar s\gamma^\mu\gamma_5 s|\eta_s(P)\rangle
   = -i f_s\,P^\mu,  & &\langle 0|\bar s\gamma^\mu\gamma_5 s|\eta_q(P),g(P)\rangle
   = -i f_{q,g}^s\,P^\mu \;.\label{deffq}
\end{eqnarray}
The decay constants associated with the $\eta$ meson, $\eta'$ meson,
and the physical glueball
are related to those associated with the $\eta_q$, $\eta_s$, and
$g$ states via the same mixing matrix
\begin{eqnarray}
\left(
\begin{array}{cc}
f_\eta^q & f_\eta^s \\
f_{\eta'}^q & f_{\eta'}^s \\
f_G^q &f_G^s
\end{array} \right)=
U(\phi,\phi_G) \left(
\begin{array}{cc}
f_q & f_q^s \\
f_s^q & f_s \\
f_g^q & f_g^s
\end{array} \right)
\;.\label{fpi}
\end{eqnarray}

Sandwiching the equations of motion for the anomalous Ward identity
\begin{eqnarray}
   \partial_\mu(\bar q\gamma^\mu\gamma_5 q) = 2im_q\,\bar q\gamma_5 q
   +\frac{\alpha_s}{4\pi}\,G_{\mu\nu}\,\widetilde{G}^{\mu\nu}\;,
   \label{eom}
\end{eqnarray}
between vacuum and $|\eta\rangle$, $|\eta'\rangle$ and $|G\rangle$, we derive
\begin{eqnarray}
\left(\begin{array}{ccc}
m_{qq}^2+(\sqrt{2}/f_q)\langle 0|q|\eta_q\rangle & m_{sq}^2+(1/f_s)\langle
0|q|\eta_q\rangle & 0\\
              m_{qs}^2+(\sqrt{2}/f_q)\langle
0|q|\eta_s\rangle &
m_{ss}^2+(1/f_s)\langle 0|q|\eta_s\rangle
&0
\\
m_{qg}^2+(\sqrt{2}/f_q)\langle 0|q|g\rangle & m_{sg}^2+(1/f_s)\langle 0|q|g\rangle &0
\end{array}\right)
=U^\dagger(\phi,\phi_G) M^2 U(\phi,\phi_G)\tilde
J\;,\label{matrix}
\end{eqnarray}
where $q=\alpha_sG{\tilde
G}/(4\pi)$ and
\begin{eqnarray}
M^2&=&\left(\begin{array}{ccc}
  m_{\eta}^2 & 0 &0\\
  0 & m_{\eta'}^2&0 \\
  0 &0 & m_G^2
\end{array} \right)\;,\;\;\;\;
\tilde J=\left(\begin{array}{ccc}
   1 & f_q^s/f_s &0\\
  f_s^q/f_q & 1 & 0 \\
  f_g^q/f_q &f_g^s/f_s & 0
\end{array} \right)\;,\label{I}
\end{eqnarray}
with the abbreviation
\begin{eqnarray}
m_{qq,qs,qg}^2 \equiv \frac{\sqrt{2}}{f_q}\langle 0|m_u\bar u i\gamma_5
u+m_d\bar d
i\gamma_5 d|\eta_q,\eta_s,g\rangle,
m_{sq,ss,sg}^2&\equiv&\frac{2}{f_s}\langle 0|m_s\bar s i\gamma_5
s|\eta_q,\eta_s,g\rangle\;. \nonumber \label{mqq}
\end{eqnarray}
Now we have six equations for many unknowns. Hence we have to reply on the large $N_c$ counting rules \cite{tHooft}
\begin{eqnarray} \label{largeNc}
&& \qquad  f_{q,s} \sim  O(\sqrt{N_c})\;, \qquad
f_g^{q,s}\sim O(1)\;, \qquad f_q^s\sim f_s^q\sim
O(1/\sqrt{N_c})\;,
\nonumber \\
&& \qquad  m_G \sim O(1), \qquad\quad\quad m_{\eta_{_8}}^2\sim O(1), \qquad m_{\eta_{_1}}^2\sim O(1)+O(1/N_c),
\nonumber\\
&& \qquad  m_{qq}^2 \sim O(1), \qquad\quad\quad m_{ss}^2\sim
O(1), \qquad \phi_G\sim O(1/\sqrt{N_c})\;,\\
&& \qquad m_{qg}^2\sim m_{sg}^2 \sim
O(1/\sqrt{N_c})\;,\qquad\quad\quad\qquad
m_{qs}^2\sim m_{sq}^2 \sim O(1/N_c)\;,   \nonumber\\
&& \qquad \langle 0|q|g\rangle \sim O(1), \qquad
\langle 0|q|\eta_q\rangle \sim \langle
0|q|\eta_s\rangle \sim O(1/\sqrt{/N_c})\;, \nonumber
\end{eqnarray}
to solve the equations step by step.

To the leading order of $1/N_c$ expansion, we shall keep the decay constants $f_q$ and $f_s$ and neglect $f_g^{q,s}$, $f_q^s$ and $f_s^q$ as they are suppressed by $1/\sqrt{N_c}$ and $1/N_c$, respectively. Likewise, we can just retain the diagonal mass terms $m_{qq}^2\approx m_\pi^2$, $m_{ss}^2\approx 2m_K^2-m_\pi^2$ and neglect other off-diagonal mass terms. It turns out that the ratio of the last two equations in the third low of Eq. (\ref{matrix}) yields
\begin{eqnarray}
\frac{c\theta (s\phi-\sqrt{2/3}c\theta\Delta_G)m_{\eta'}^2-s\theta
(c\phi+\sqrt{2/3}s\theta\Delta_G)^2m_\eta^2 -\sqrt{2/3} c\phi_G
m_G^2}{c\theta (c\phi-\sqrt{1/3}c\theta\Delta_G)m_{\eta'}^2+s\theta
(s\phi-\sqrt{1/3}s\theta\Delta_G)^2m_\eta^2 -\sqrt{1/3}c\phi_G
m_G^2}=\frac{\sqrt{2}f_s}{f_q}\label{rg}
\end{eqnarray}
where $c\phi$ ($s\phi$) is the shorthand notation for $\cos\phi$
($\sin\phi$) and similarly for others. {\it This simple equation tells us that the pseudoscalar glueball mass $m_G$ can be determined provided that the mixing angle $\phi_G$ and the ratio $f_s/f_q$ are known.} Note that the $\phi_G$ dependence appears at
order of $\Delta_G\approx \phi_G^2$ for small $\phi_G$. So the solution for $m_G$ is stable against the most
uncertain input $\phi_G$.

The mixing angles $\phi$ and $\phi_G$
have been measured recently from the $\phi \to \gamma\eta,
\gamma\eta'$ decays by KLOE \cite{KLOE}. Using the decay constants $f_q=(1\pm 0.01)f_\pi$ and
$f_s=(1.4\pm 0.014)f_\pi$  as inputs \cite{TF00}, KLOE obtained
the angles $\phi=(39.7\pm 0.7)^\circ$ and $\phi_G=(22\pm 3)^\circ$
inferred from the relevant data. In
\cite{EN07} the data of $P\to \gamma V$ and $V\to \gamma P$ were
first considered and the
fit gave the outcomes $\phi=(41.4\pm 1.3)^\circ$ and $\phi_G=(12\pm
13)^\circ$. Without precise inputs of $f_q$ and $f_s$ it
is not unexpected to get a wide range for $\phi_G$. Since
$\phi_G$ has a wide range, the results $f_q=(1.05\pm 0.03)f_\pi$ and
$f_s= (1.57\pm 0.28)f_\pi$ also have larger errors. Using the central values of
$f_s/f_q$ and $\phi_G$ from \cite{KLOE,EN07} as inputs, we
derive the pseudoscalar glueball mass from Eq.~(\ref{rg}) to be
\begin{equation} \label{mG}
m_G=(1.4\pm0.1)~{\rm GeV}.
\end{equation}
The proximity of the predicted $m_G$ to the mass of
$\eta(1405)$ and other properties of $\eta(1405)$ make it a very
strong candidate for the pseudoscalar glueball.

Our next task is to check the stability and robustness of our prediction when higher
order effects in $1/N_c$ are included. We first turn on the decay constants $f_g^{q,s}$ and $f_q^s$, $f_s^q$. Assuming flavor-independent couplings between the glueball $g$
and the pseudoscalar $u\bar{u}, d\bar{d}$ and $s\bar{s}$ states, we then have the relations
\begin{equation} \label{fg}
f_g^q = \sqrt{2} f_g^s, \qquad f_s^q =f_q^s.
\end{equation}
It turns out that the above simple formula Eq. (\ref{rg}) still holds even after keeping the OZI-correcting decay constants, as long as they obey Eq.~(\ref{fg}).

\begin{table}[t]
\tbl{Solutions for the input of $\langle
0|q|\eta_q\rangle=0.050$ GeV$^3$  and $f_s$ being fixed to trade
for $m_{sg}^2$ as a free parameter. The upper
(lower) table is for $\phi_G=22^\circ$ ($\phi_G=12^\circ$), $r\equiv f_g^s/f_s$ and $R
\equiv f_q^s/f_s$.}
{\begin{tabular}{c ccc c c c} \toprule
$f_s$ & $r$ & $R$ & $m_{sg}^2$\,(GeV$^2$) & $m_G$\,(GeV) & $\langle
0|q|\eta_s\rangle$\,(GeV$^3$)
& $\langle 0|q|g\rangle$\,(GeV$^3$) \\
\colrule
$1.24 f_\pi$ & $0.22$ & $-0.001$ & $-0.009$  & $1.60$ & 0.028 & $0.036$   \\
$1.26 f_\pi$ &
$0.22$ & $-0.003$ & $0.004$  & $1.47$ & 0.028 & 0.036   \\
$1.28 f_\pi$ &
$0.23$ & $-0.005$ & $0.016$  & $1.34$ & 0.028 & 0.038   \\
$1.30 f_\pi$ &
$0.24$ & $-0.007$ & $0.029$  & $1.21$ & 0.028 & 0.040   \\
\colrule
$1.24 f_\pi$ &
$0.12$ & $0.001$ & $-0.054$  & $2.15$ & 0.027 & $0.030$   \\
$1.26 f_\pi$ &
$0.13$ & $-0.001$ & $-0.029$ & $1.84$ & 0.027 & 0.031   \\
$1.28 f_\pi$ &
$0.15$ & $-0.003$ & $-0.005$  & $1.52$ & 0.027 & 0.034   \\
$1.30 f_\pi$ &
$0.24$ & $-0.005$ & $0.018$  & $1.16$ & 0.028 & 0.045   \\
\botrule
\end{tabular}} \label{tab:2}
\end{table}

Finally we turn on the mass term $m_{sg}^2$ and neglect $m_{qg}^2$, $m_{qs}^2$ and $m_{sq}^2$. It is justified to do so because $m_{qg}^2$ is proportional to the light $u/d$ quark mass, while the last two mass terms are $1/\sqrt{N_c}$ suppressed relative to $m_{sg}^2$. To explore the impact of $m_{sg}^2$ on our solutions, we add $f_s$ as an input so that $m_{sg}^2$ can be
introduced as an unknown.  The results
for the various inputs of $f_s=(1.24$-$1.30) f_\pi$,
$\phi_G=22^\circ$ and $12^\circ$, and $\langle 0|q|\eta_q\rangle=0.050$ GeV$^3$ are listed in
Table~\ref{tab:2}. We see that $m_{sg}^2$ and $m_G$ do depend on
$f_s$ sensitively. In some cases, we have $m_G$ as large as 1.84 GeV
and 2.15 GeV, for which $m_{sg}^2$ are negative and large. We cannot
discard these solutions of $m_{sg}^2$ {\it a priori}, but they are
not favored due to their negative values. This issue can be sorted
out, when lattice calculations of $m_{sg}^2$ with dynamical fermions
are available.
Therefore, if excluding the solutions with large and negative
$m_{sg}^2$,
the range $(1.4 \pm 0.1)$ GeV of the pseudoscalar glueball mass
obtained in Eq. (\ref{mG}) will be more or less respected.

One may feel rather uncomfortable with our solution for $m_G$ as both lattice QCD and QCD sum rules indicate a pseudoscalar glueball heavier than the scalar one.
The point is that lattice calculations so far were
performed under the quenched approximation without the fermion
determinants. It is believed that dynamical fermions will have a
significant effect in the pseudoscalar channel, because they raise
the singlet would-be-Goldstone boson mass from that of the pion to
$\eta$ and $\eta'$. Indeed, it has been argued that the pseudoscalar
glueball mass in full QCD is substantially lower than that in the
quenched approximation \cite{GG97}. In view of the fact that the
topological susceptibility is large (of order $10^{-3}\,{\rm GeV}^4$) in
the quenched approximation, and yet is of order $10^{-5}\,{\rm GeV}^4$ in full QCD and zero in the chiral limit, it is conceivable that full QCD has a large
effect on the pseudoscalar glueball as it does on $\eta$ and $\eta'$.

According to our
analysis, the $\eta(1405)\to \gamma\gamma$ decay width is
0.6-3 keV, and the leptonic decays $\eta(1405)\to\ell^+\ell^-$ are
very small \cite{CLL}. Both predictions can be confronted with future experiments. There may not exist a unique feature which tells a glueball apart
from a quark-antiquark state. We need to combine information from
$J/\psi$ radiative decays, hadronic decays, as well as
$\gamma\gamma$ and leptonic decays as advocated in \cite{LLI89}.

\section{Conclusions}
We have employed two simple and robust results to constrain the mixing matrix of the isosinglet scalar mesons
$f_0(1710)$, $f_0(1500)$, $f_0(1370)$: (i) empiric SU(3) symmetry in the scalar sector
above 1 GeV, and (ii) the scalar glueball mass at 1710
MeV in the quenched approximation. In the SU(3) symmetry limit,
$f_0(1500)$ becomes a pure SU(3) octet and is degenerate with
$a_0(1450)$, while $f_0(1370)$ is mainly an SU(3) singlet with a
slight mixing with the scalar glueball which is the primary
component of $f_0(1710)$. These features remain essentially
unchanged even when SU(3) breaking is taken into account.

From the analysis of the $\eta-\eta'-G$ mixing together
with the inputs from the
recent KLOE experiment, we find a solution for the pseudoscalar
glueball mass around $(1.4\pm 0.1)$ GeV, suggesting that $\eta(1405)$ is indeed a leading candidate for the
pseudoscalar glueball.  It is thus urgent to have a full QCD lattice calculation of pseudoscalar glueball masses.

\section*{Acknowledgments}
I'm very grateful to  Chun-Khiang Chua, Hsiang-nan Li and Keh-Fei Liu  for the fruitful collaboration
on glueballs and to K K Phua for organizing this wonderful conference.

\end{document}